\documentclass[preprint2]{aastex}
\input psfig.sty

\def\spose#1{\hbox to 0pt{#1\hss}}
\def\simlt{\mathrel{\spose{\lower 3pt\hbox{$\mathchar"218$}}
     \raise 2.0pt\hbox{$\mathchar"13C$}}}
\def\simgt{\mathrel{\spose{\lower 3pt\hbox{$\mathchar"218$}}
     \raise 2.0pt\hbox{$\mathchar"13E$}}}

\slugcomment{Submitted to AJ}

\shorttitle{The Clustering of AGN and Galaxies}
\shortauthors{Brown et al.}

\begin{document}


\title{The Clustering of AGN and Galaxies at Intermediate Redshift}


\author{M. J. I. Brown}
\affil{National Optical Astronomy Observatory, P.O. Box 26732, 
950 North Cherry Avenue, Tucson, AZ 85726, USA}
\email{mbrown@noao.edu}

\author{B. J. Boyle}
\affil{Anglo-Australian Observatory, P.O. Box 296, Epping, NSW 1710, Australia}

\and

\author{R. L. Webster}
\affil{School of Physics, University of Melbourne, Parkville, Victoria 3010, Australia}

\begin{abstract}
Galaxies in the environments of 69 $0.2<z<0.7$ $UBR$ selected AGN have been 
imaged to $B_J\sim 23.5$. By applying photometric redshifts 
and color selection criteria to the galaxy catalogue, 
the AGN-galaxy cross-correlation function has 
been measured as a function of galaxy type. The spatial cross-correlation
of AGN with red (early-type) galaxies is comparable to the 
autocorrelation function of elliptical galaxies at low redshift. 
In contrast, the cross-correlation of AGN with blue (late-type) 
galaxies is weak and has been detected with low significance. 
As blue galaxies dominate $B_J \sim 23.5$ galaxy catalogues, 
the cross-correlation of $UBR$ selected AGN with all galaxies is weak at 
intermediate redshifts. 
\end{abstract}


\section{Introduction}

The association of Active Galactic Nuclei hosts with other galaxies
is well established \citep{bah69,yee87,hal98,lau95,smi95,smi00} and 
provides constraints on the models for the formation and fueling of AGN. 
In addition, as QSOs can be used to trace large-scale-structure at 
$z\gg 1$ \citep{boy99}, estimates of the QSO environments 
are required to estimate their bias with respect to galaxies 
which are used to trace large-scale-structure at $z<1$. 

Several previous studies of AGN environments are summarized in
Table~\ref{table:qsogal}. Radio-loud AGN are typically found in 
environments comparable to galaxy clusters \citep{yee87,hal98,wol00}
while $z<0.3$ radio-quiet AGN appear to be associated with poorer environments
similar to field galaxies \citep{lau95, smi95, der98}. While the 
$z>0.3$ radio-quiet AGN-galaxy cross-correlation function 
is comparable to the clustering of field galaxies \citep{smi00}, 
it has only been detected with
$\sim 2\sigma$ significance \citep{ell91, tep99, smi00} 
and no detailed information is available on the strength or evolution of
galaxy clustering around radio-quiet AGN at these redshifts.

Previous studies of AGN environments have typically used galaxy catalogues
derived from single band imaging. Galaxy evolution and $k-$corrections
result in a changing morphological mix of galaxies as a function of
redshift and depth. As galaxy clustering is correlated with 
morphology and color, the clustering properties of galaxies
are also a function of redshift and the observer's bandpass. It is therefore 
plausible that estimates of AGN environments at $z>0.3$ have been biased
by the changing properties of the galaxy catalogues selected from 
single band imaging. By using color selection, it is possible to 
select the same galaxy type as a function of redshift. By measuring
the cross-correlation function of AGN with early and late-type galaxies,
it is possible to determine if radio-quiet AGN are in unusual environments.

\section{Data}

\subsection{The Galaxy Sample}

The galaxy sample has been previously used by \cite{bro00} to measure 
the clustering of galaxies as a function of color and a more detailed
description of the catalogue is provided by \cite{bro00} and \cite{phd}. 
The image data consists of $5^\circ\times 5^\circ$ images of the 
South Galactic Pole (SGP) and UK Schmidt field 855 (F855). The images
were produced by stacking SuperCOSMOS scans of UK Schmidt photographic
plates in $U$, $B_J$, $R$ and $I$ bands. 
Object detection, instrumental photometry and faint object 
star-galaxy classifications were determined with
SExtractor \cite{ber96}. Photometric calibration of the data
was determined with CCD images and published photometry. To prevent
dust extinction introducing spurious large-scale-structure, magnitude
estimates are corrected with the extinction estimates of \cite{sch98}.
The final galaxy catalogues are complete
to $B_J\sim 23.5$ and contain $\sim 2\times 10^5$ galaxies per field. 

\subsection{The AGN Sample}
\label{sec:agnsample}

\begin{figure}
\centerline{\psfig{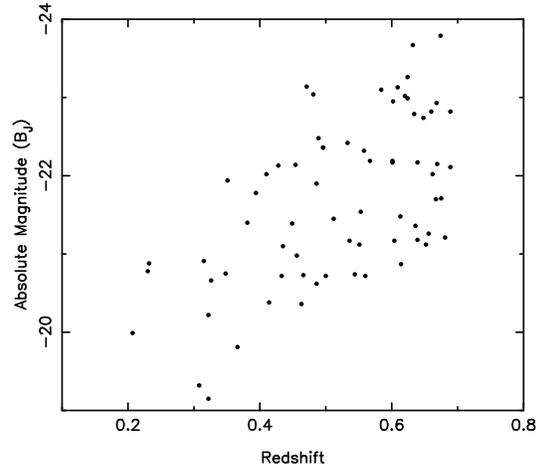}}
\figcaption[MJIBrown.fig1.ps]
{The absolute magnitudes and redshifts of the $0.2<z<0.7$ AGN
selected from the \cite{laf99} catalogue. The absolute
magnitudes have been determined with $\Omega=1$, 
$H_0=100 {\rm km s}^{-1} {\rm Mpc}^{-1}$ and $k-$corrections approximated
by $k(z)=-0.4z$.
\label{fig:qmagz}}
\end{figure}

The AGN sample consists of 69 $0.2<z<0.7$ $UBR$ selected broad 
emission line AGN from \cite{laf99}. The survey area is the SGP field
and while the catalogue does not have homogeneous sky coverage, it is 
not strongly concentrated in any one part of the field. As the F855 field
only contains 7 $0.2<z<0.7$ AGN with published positions 
\citep{ver00}, the F855 field has not been used to measure AGN-galaxy 
clustering. AGN positions
were determined by selecting $B_J<23.5$ objects in the stacked SuperCOSMOS
scans within $5^{\prime\prime}$ of the published positions. 
The resulting catalogue of AGN contained 69 $B_J<21$ 
objects with $U-B_J<0.1$ colors. The $B_J$ 
absolute magnitudes and redshifts of the AGN are plotted in 
Figure \ref{fig:qmagz}. The catalogue contains 36 QSOs 
($M_{B_J}<-21.5+5{\rm log} h$ where $h \times 100 {\rm kms}^{-1} {\rm Mpc}^{-1} \equiv H_0$) 
and 33 Seyfert 1 galaxies. Five of the 
AGN are within $20^{\prime\prime}$ of $1.4~{\rm GHz}$ sources detected 
by the NVSS \citep{con98} but all have fluxes less than 
$60~{\rm mJy}$ and $1.4~{\rm GHz}$ luminosities less than 
$10^{26} {\rm W Hz}^{-1}$. 

\subsection{Photometric Redshifts}

\begin{figure}
\centerline{\psfig{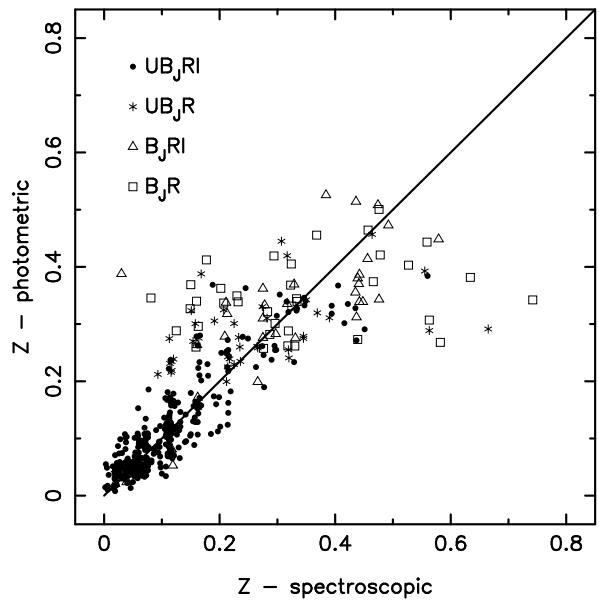}}
\figcaption[MJIBrown.fig2.ps]
{Comparison of photometric and spectroscopic redshifts of
galaxies. The accuracy of the redshift estimates derived with 3 or
more colors is significantly better than redshifts derived with
$B_JR$ photometry.
\label{fig:photoz}}
\end{figure}

\begin{figure}
\centerline{\psfig{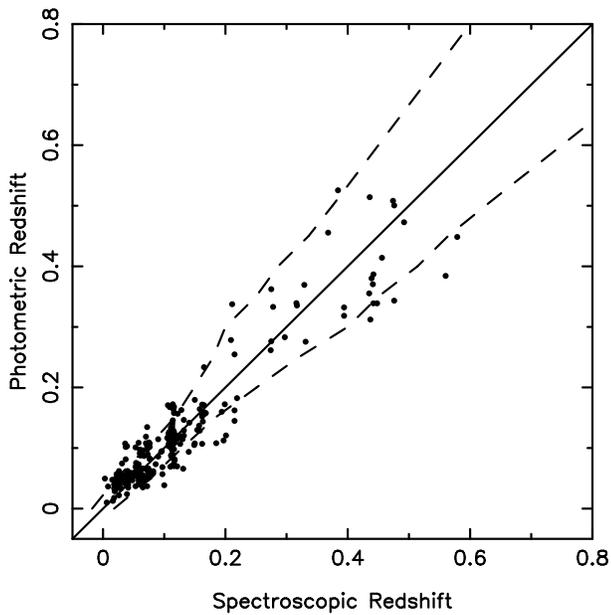}}
\figcaption[MJIBrown.fig3.ps]
{Comparison of photometric and spectroscopic redshifts for
galaxies with colors redder than non-evolving
Sbc galaxies. The dashed lines show the  $\pm 1\sigma$ error estimates of
the photometric redshifts. While some of the galaxies are only
detected in  $B_J$ and $R$, the correlation between photometric
and spectroscopic redshift is still good.
\label{fig:Sbcphotoz}}
\end{figure}

Approximately $700$ $0<z<0.8$ galaxy redshifts in the SGP and F855 fields are
available from the NED database. Spectroscopic redshifts are available
for galaxies as faint as $B_J\sim 24$ so it possible to use 
multicolor photometry and spectroscopic redshifts of galaxies 
in the SGP and F855 fields to calibrate the photometric redshifts. 
The relationship between the multicolor photometry and the
galaxy redshift was determined by fitting quadratic functions to
the data \citep{con95}. The relationship was determined for $UB_JRI$,
$UB_JR$, $B_JRI$ and $B_JR$ photometry as only a small
fraction of the catalogue is detected in all 4 bands.
Figure \ref{fig:photoz} shows a comparison
photometric redshifts and the spectroscopic redshifts for
galaxies in the two fields. While $B_JR$ photometric redshifts are
poorer than photometric redshifts derived in 3 or more bands, they
do place useful constraints on the redshifts. 
Error estimates for the photometric redshifts have been
determined by the measuring the rms of the residuals as a function
of photometric redshift and color. The accuracy of the photometric redshifts
is a function of galaxy color and, as shown in Figure~\ref{fig:Sbcphotoz},
red galaxies have comparatively small errors. 

\subsection{Color Selection}

A significant bias present in most studies of the AGN environment is that 
they have relied on small samples or single band imaging.
Galaxy evolution and $k-$corrections result in a changing morphological
mix of galaxies as a function of limiting magnitude with single band imaging. 
Galaxy catalogues in bands bluer than $R$ are dominated by 
weakly clustered (blue) late-type galaxies at magnitudes 
fainter than $B_J\sim 22$ \citep{efs91}. 

Multicolor imaging provides significant advantages for the 
study of the AGN host environment. Colors and photometric 
redshifts can be used to select particular galaxy types
at faint magnitudes. As galaxy clustering is strongly correlated with color 
\citep{bro00} and morphology \citep{dav76, lov95}, 
studying the cross-correlation of AGN with particular galaxy types 
should help determine if AGN hosts are in unusual galaxy environments.

\begin{figure}
\centerline{\psfig{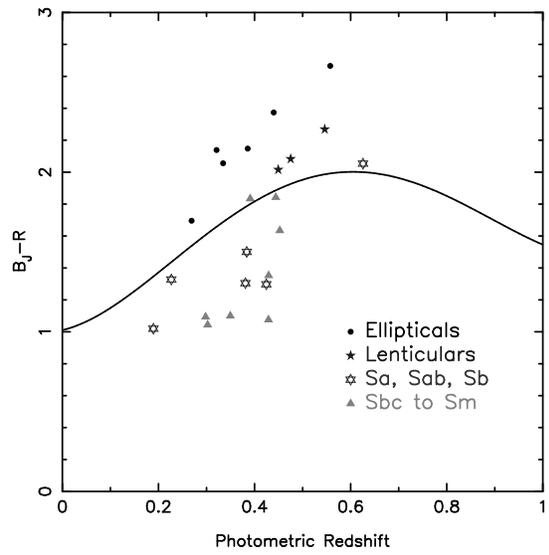}}
\figcaption[MJIBrown.fig4.ps]
{The colors and photometric redshifts of galaxies with
morphological classifications from \cite{abr96}
and \cite{sma97}. The curved line is an estimate of
the color of a non-evolving Sbc determined with a polynomial fit to the
$k-$corrections from \cite{col80}. Most early-type
galaxies are redder than the non-evolving Sbc.
\label{fig:cwwSbc}}
\end{figure}

The color criteria applied to the galaxy catalogue
select galaxies redder or bluer than a non-evolving Sbc in $B_J-R$ where
the color as a function of redshift is determined with 
$k-$corrections from \cite{col80}.
The use of $B_J-R$ rather than shallower $U-B_J$ or $R-I$ allows the 
selection of $B_J=23.5$ early-type galaxies at $z\sim 0.5$. 
As the $B_J-R$ selection criteria are a function of redshift, photometric
redshifts are used to determine the correct value of $B_J-R$ when 
selecting galaxies. For the remainder of this paper, the 
early and late subsamples will refer to galaxies redder and 
bluer than the $B_J-R$ selection criteria. 
Figure \ref{fig:cwwSbc}, a plot of the colors and photometric
redshifts of galaxies in the SGP with HST morphologies 
\citep{abr96, sma97}, shows that the color selection
criteria are capable of selecting different types of 
galaxies. 

\section{The Angular and Spatial Correlation Functions}

The angular correlation function estimates the fractional excess of object
pairs at a given angular separation compared with what would be expected for
a random distribution of objects. The estimators of the angular correlation
function used in this paper require random object catalogues which are 
produced by making copies of the galaxy catalogue and randomly repositioning the 
copied galaxies across the field. 

The angular autocorrelation function of galaxies is 
estimated with 
\begin{equation}
\hat\omega(\theta)=\frac{DD-2DR+RR}{RR}
\end{equation} 
\citep{lan93} where $DD$, $DR$ and $RR$ are the number of 
galaxy-galaxy, galaxy-random and random-random pairs at angular 
separation $\theta\pm\delta\theta$. The \cite{lan93} estimator
is only applicable to the autocorrelation function so the 
AGN-galaxy (and early-late) cross-correlation function 
is determined with 
\begin{equation}
\hat\omega(\theta)=\frac{AG}{AR} - 1
\end{equation} 
where $AG$ and $AR$ are the number of AGN-galaxy and AGN-random 
pairs at angular separation $\theta\pm\delta\theta$.
The random errors of $AR$, $DR$ and $RR$ are reduced by taking the average 
of multiple estimates of each parameter where each estimate has been determined
with a different random object catalogue. 
Both estimators of the angular correlation function satisfy the 
integral constraint,
\begin{equation}
\int \int \hat \omega (\theta) \delta \Omega_1 \delta \Omega_2 \simeq 0
\end{equation} 
\citep{gro77}, resulting in an underestimate of the 
angular correlation function. To remove this bias from 
the correlation function, the term
\begin{equation}
\omega(\theta)_\Omega =
\frac{1}{\Omega^2} \int \int \omega (\theta) \delta \Omega_1 \delta \Omega_2
\end{equation}
is added to the estimate of the correlation function. The term
$\omega (\theta)_\Omega$ does require an assumption of the form of the
correlation function to correctly estimate the value of correlation
function. 

Galaxy angular correlation functions are typically approximated by 
power-laws where
\begin{equation}
\omega(\theta)=A\theta^{1-\gamma}
\end{equation}
where $A$ is a constant and $\gamma\sim1.7$. The corresponding 
spatial correlation function is given by 
\begin{equation}
\xi(r,z)=(r/r_0)^{-\gamma}(1+z)^{-(3+\epsilon)}
\end{equation}
where $r$ is the separation of the galaxies in physical coordinates
and $r_0$ and $\epsilon$ are constants.
For $\epsilon=0$ and $\epsilon=\gamma-3$, the clustering properties are 
fixed in physical and comoving coordinates respectively. 
The relationship between the angular and spatial correlation functions
is given by Limber's equation \citep{lim54}. If $\xi (r,z)\sim 0$ when 
$r\simgt 0.1z$, then Limber's equation is given by 
\begin{eqnarray}
\omega(\theta) = & \int \frac{dN_1}{dz} 
\left[ \int \xi (r(\theta),z) \frac{dN_2}{dz^\prime} dz^\prime \right] dz 
\nonumber \\
& \left/ \int \frac{dN_1}{dz} dz  \int  \frac{dN_2}{dz} dz \right. 
\end{eqnarray}
\citep{phi78} where $\frac{dN_1}{dz}$ and $\frac{dN_2}{dz}$ are the redshift 
distributions of each set of objects 
(e.g. the early and late subsamples) and 
$r(\theta)$ is the distance between two objects at 
$z$ and $z^\prime$ separated by angle $\theta$ on the sky.
For the AGN-galaxy correlation function, the redshifts of each AGN are known
so Limber's equation can be written as 
\begin{eqnarray}
\omega(\theta)= & \sum_i^n \left[
\int \xi_{ag} (r(\theta),z) \frac{dN_g}{dz} dz 
\right] \nonumber \\
& \left/  \sum_i^n \left[ \int \frac{dN_g}{dz} dz 
\right] \right. 
\end{eqnarray}
where $n$ is the number of AGN, $\frac{dN_g}{dz}$ is the number of 
galaxies per unit redshift and $r(\theta)$ is the distance in physical 
coordinates between an AGN at redshift $z_i$ and a galaxy at 
redshift $z$ separated by angle $\theta$ on the sky.  

\section{The Clustering of Faint Galaxies}

\begin{figure}
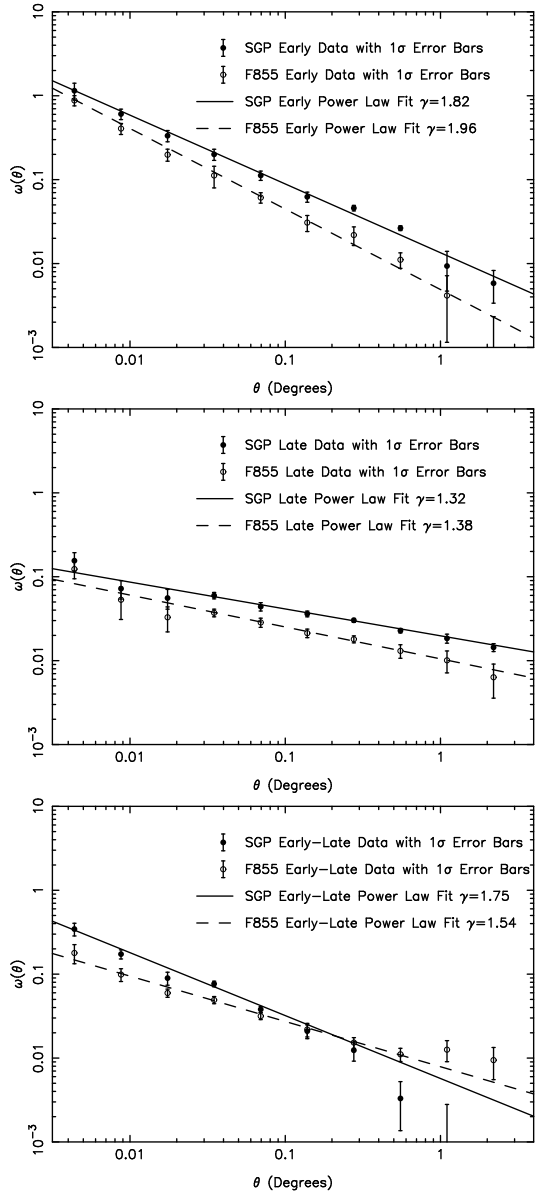

\centerline{\psfig{file=MJIBrown.fig5a.ps,angle=270,width=7.0cm}}
\centerline{\psfig{file=MJIBrown.fig5b.ps,angle=270,width=7.0cm}}
\centerline{\psfig{file=MJIBrown.fig5c.ps,angle=270,width=7.0cm}}
\figcaption[MJIBrown.fig5a.ps MJIBrown.fig5b.ps MJIBrown.fig5c.ps]
{The $18.0<B_J<22.5$ galaxy angular correlation functions for the
SGP and F855 fields. The autocorrelation function of early subsample
galaxies is significantly stronger than the autocorrelation function
of late subsample galaxies. Power laws fitted to the $<0.3^\circ$
data are good approximations to the observed clustering on most 
angular scales. 
\label{fig:galgal}}
\end{figure}

Estimates of the clustering of $z\sim 0.5$ galaxies are required
if the environment of AGN hosts is to be compared to the ``normal''
galaxy environment. To make the comparison valid, the same galaxy 
selection criteria are applied to the study of galaxy-galaxy clustering
as AGN-galaxy clustering. 
The early and late subsample autocorrelation functions and the early-late
cross-correlation functions have been determined in the SGP and 
F855 fields. Plots of the $18.0<B_J<22.5$ angular correlation functions
are shown in Figure~\ref{fig:galgal}.
The amplitudes and values of the $\gamma$ as a function
of magnitude are summarized in Tables \ref{table:ee} to \ref{table:el}.
The value of $\gamma$ and the amplitude of the clustering strongly depend
on color \citep{bro00}. The observed clustering in the SGP is consistently
stronger than the observed clustering in F855 at bright magnitudes. 
However, at fainter magnitudes the difference between the clustering 
properties in the two fields is significantly reduced and the late 
subsample autocorrelation functions are comparable. 
At faint magnitudes, the estimates of $\gamma$ for the early-late 
cross-correlation function in the SGP and F855 fields differ by $\sim 2 \sigma$. 
It is possible that this is due to the cross-correlation function estimator
having larger errors than predicted by the Poisson estimate
\citep{lan93}. Also, at faint magnitudes in the F855 field, the 
early-late cross-correlation function is comparable to 
the expected variations in the galaxy number counts introduced by errors 
in the dust extinction estimates.

To determine the spatial correlation function, an estimate of
the redshift distribution is required. Estimates of the 
redshift distribution derived from luminosity functions, 
$k-$corrections and evolution models are subject to uncertainties 
as a range of models can reproduce the observed number counts. 
Models which assume a shape for the redshift distribution 
\citep{bau93} are useful for single band imaging
of low redshift galaxies but are not as effective for multicolor selected
samples at higher redshifts where the selection criteria and $k-$corrections
skew the redshift distribution.

\begin{figure}
\centerline{\psfig{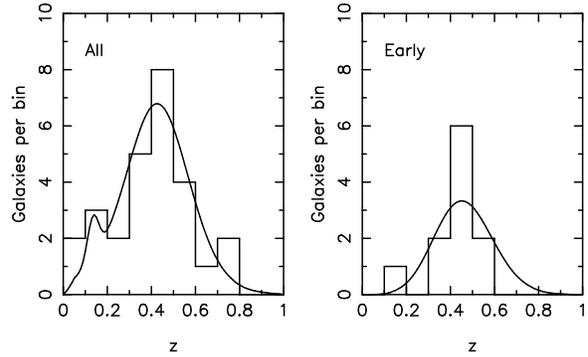}}
\figcaption[MJIBrown.fig6.ps]
{A comparison of the observed and model
redshift distributions of galaxies from \cite{gla95}
which were detected in the stacked scans. The left panel shows
all galaxy types while the right panel shows galaxies with colors
redder than Sbc galaxies. While limited by 
the small number of galaxies available, the models are similar
to the observed redshift distributions with both peaking at 
similar redshifts. 
\label{fig:glazephotoz}}
\end{figure}

Photometric redshifts contain information on the redshift distribution
but assume that galaxies with the same colors and magnitudes are at 
the same redshifts. If the errors of the photometric redshifts are
dominated by the redshift distribution of galaxies with the same 
multicolor photometry, it should be possible to derive
a redshift distribution using the measured errors of the photometric
redshifts. The redshift distribution has been estimated by 
smoothing the photometric redshift distribution with a Gaussian where the 
$\sigma$ of the Gaussian is given by the rms of the errors of the 
redshift estimates. At $z<0.05$ the 
redshift distribution has been multiplied by $z/0.05$ to prevent
an infinite density of galaxies at $z=0$.
Figure \ref{fig:glazephotoz} shows the observed 
redshift distribution of galaxies from \cite{gla95} 
detected in the stacked scans and models derived from 
their photometric redshifts. There 
is reasonable agreement between the measured and model distributions, 
though the galaxy number counts are limited. While 
the exact redshift distributions have not been determined, it is 
unlikely that large errors dominate the model redshift distributions. 

\begin{figure}
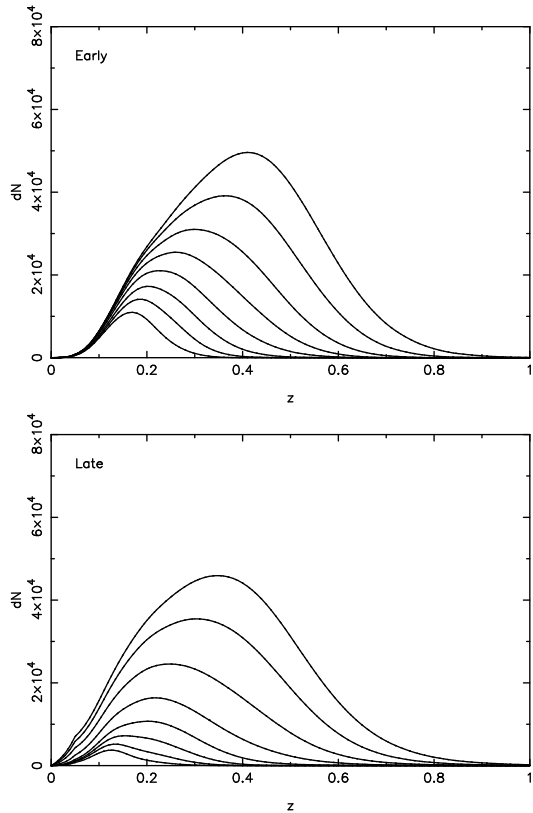

\centerline{\psfig{file=MJIBrown.fig7a.ps,angle=270,width=7.0cm}}
\centerline{\psfig{file=MJIBrown.fig7b.ps,angle=270,width=7.0cm}}
\figcaption[MJIBrown.fig7a.ps MJIBrown.fig7b.ps]
{The estimated redshift distribution of the early and late subsamples in the
SGP. The curves range from $18.0<B_J<20.0$ to $18.0<B_J<23.5$ in 
half magnitude steps. The large redshift errors for blue galaxies results in the
late subsample redshift distribution being significantly 
broader than the early subsample redshift distribution.  At magnitudes fainter
than $B_J=22.5$ photometric redshifts are not available for all late
subsample galaxies and this may produce errors in the model redshift distribution.
\label{fig:dnearly}}
\end{figure}

The estimated redshift distributions of the early and late subsamples
are plotted in Figure \ref{fig:dnearly}. The photometric redshifts
of blue galaxies have large error estimates and this results
in the late subsample redshift distribution being significantly 
broader than the early subsample redshift distribution.  
The width of the distribution is consistent with late-type galaxies having 
smaller $k-$corrections and a higher fraction of dwarf galaxies
than early-type galaxies. It should be noted that while the photometric
redshifts are complete to $B_J=23.5$ for the early subsample, photometric
redshifts for $B_J>22.5$ late subsample galaxies are incomplete and this
may slightly skew the redshift distribution. 

\begin{figure}
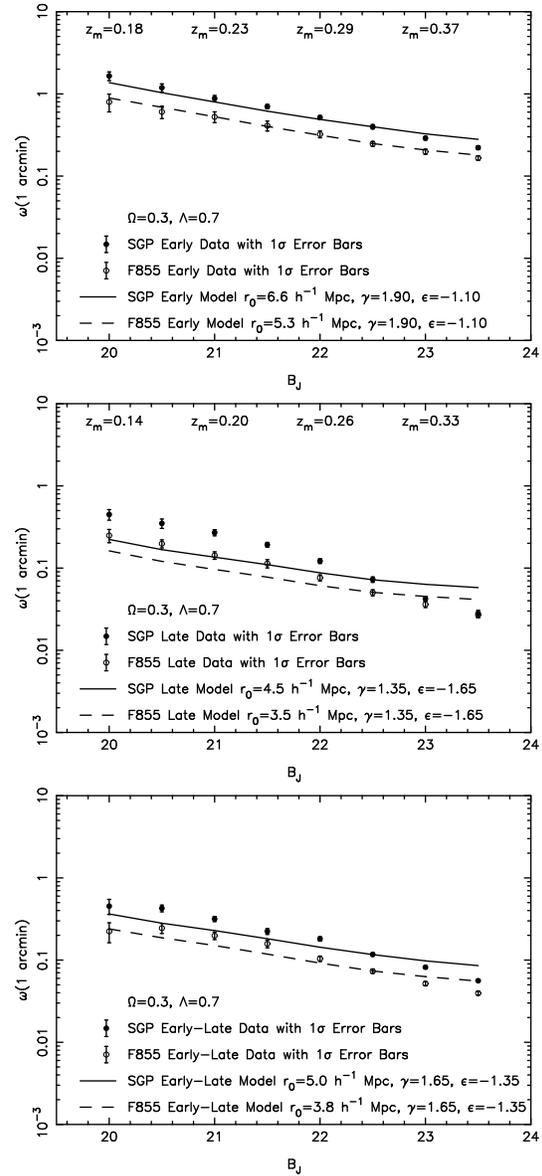

\centerline{\psfig{file=MJIBrown.fig8a.ps,angle=270,width=7.0cm}}
\centerline{\psfig{file=MJIBrown.fig8b.ps,angle=270,width=7.0cm}}
\centerline{\psfig{file=MJIBrown.fig8c.ps,angle=270,width=7.0cm}}
\figcaption[MJIBrown.fig8a.ps MJIBrown.fig8b.ps MJIBrown.fig8c.ps]
{The amplitudes of the early and late autocorrelation functions and 
the early-late cross-correlation function. 
Models fitted to the $B_J=22.5$ data points with clustering
fixed in comoving coordinates are shown. The median redshift of the 
sample as a function of depth is listed at the top of the plot.
The difference between the observed clustering in the two fields 
generally decreases as a function of magnitude. 
\label{fig:ggampmag}}
\end{figure}

The amplitudes of the early and late autocorrelation functions and 
early-late cross-correlation function have been plotted in 
Figure \ref{fig:ggampmag}. To allow the comparison of the 
amplitude at different magnitudes, $\gamma$ has been fixed for 
all magnitudes to the average value of $\gamma$ for the SGP and F855 
$18.0<B_J<23.5$ correlation functions. Models with the clustering 
fixed in comoving coordinates have been fitted to the $18.0<B_J<22.5$
magnitude range for the SGP and F855 fields. 
The models have been fitted to this magnitude range as it includes 
a large number of galaxy-galaxy pairs while avoiding systematic errors
which could be present near the magnitude limits of the data. 
Additional estimates of $r_0$ with clustering fixed in comoving and physical 
coordinates are provided in Table~\ref{table:r0}.

Figure~\ref{fig:ggampmag} shows the clustering in the SGP is 
consistently stronger than the clustering in F855 for 
$18.0<B_J<22.5$ galaxies. Before concluding 
that large-scale-structure is responsible for the observed 
difference between the clustering in the two fields, it is useful
to exclude the most plausible sources of error. 
Systematic errors in the $B_J-R$ colors of $\sim 0.2$ magnitudes 
in the two fields only slightly alter the observed clustering in 
each field and are inconsistent with galaxy number counts, color-color 
diagrams and published photometry. Large uniform errors in the photometry in 
all 4 bands can bring the clustering at bright magnitudes into
agreement but are also inconsistent with galaxy number counts and 
published photometry. The \cite{sch98} dust maps estimate
$E(B-V)\sim 0.05$ in the F855 field and errors in the dust 
extinction estimates could introduce systematic errors. 
However, errors in the dust extinction estimates should produce 
the largest discrepencies at faint magnitudes where the amplitude
of the angular correlation function is small rather than at bright 
magnitudes. 

It is therefore not implausible that the difference between 
the two fields is due to 
large-scale-structure. The distribution of galaxy clusters
in both fields shows evidence of structures comparable to the 
field-of-view \citep{bro00, phd} and the SGP may 
contain several ``sheets'' of galaxies \citep{bro90} though it is 
uncertain if this is particularly unusual \citep{kai91}. 
The effects of large-scale-structure on estimates of the correlation function
should decrease with increasing survey volume and this is consistent 
with the observed difference between the clustering in the two fields decreasing 
with increasing depth. While it is unexpected to see weakly clustered
late subsample galaxies showing the effects of large-scale-structure, 
the discrepancy between the two fields disappears for bluer galaxies 
selected with $U-B_J<0.4$
\citep{bro00}. As it is uncertain which field is 
more representative of the Universe, the discussion of
AGN-galaxy clustering in the SGP assumes that estimates of $r_0$ could have 
systematic errors comparable to the difference between the estimates of 
$r_0$ in the SGP and F855 ($\sim 1 h^{-1} {\rm Mpc}$). 

Models for the clustering of early subsample galaxies
in each field are a reasonable approximation to the observed 
clustering and the estimates of the spatial correlation function 
are comparable to the clustering of early-type galaxies in the 
local Universe \citep{lov95,guz97}. In contrast, the 
observed amplitude of the late subsample autocorrelation function
is an order of magnitude weaker and the models are a poor fit to
the data. At $B_J>22$ the amplitude of the correlation function 
rapidly decreases due to the increasing number of weakly clustered
faint blue galaxies \citep{efs91,bro00}.
The amplitude of the early-late cross-correlation function also decreases 
at faint magnitudes though models of the spatial correlation
function are better approximations to the observed clustering. 
While the values of $r_0$ for the early-late cross-correlation function 
are similar to those for the late subsample autocorrelation
function, the larger value of $\gamma$ results in 
significantly stronger clustering on scales of $\simlt 1 h^{-1} {\rm Mpc}$.

\section{AGN-Galaxy Clustering}
\label{sec:agngal}

Estimates of the cross-correlation of AGN 
with the early and late subsamples are summarized in 
Table \ref{table:qe}. The AGN-early cross-correlation is significantly 
stronger on scales $\simgt 1^\prime$ while on smaller scales the 
signal-to-noise is poor. Despite the strength of the 
AGN-early cross-correlation, inspection of 
Table \ref{table:qe} shows most AGN-galaxy pairs are AGN-late pairs
and the cross-correlation of AGN with all galaxies will be similar
to the AGN-late cross-correlation function. 

\begin{figure}
\centerline{\psfig{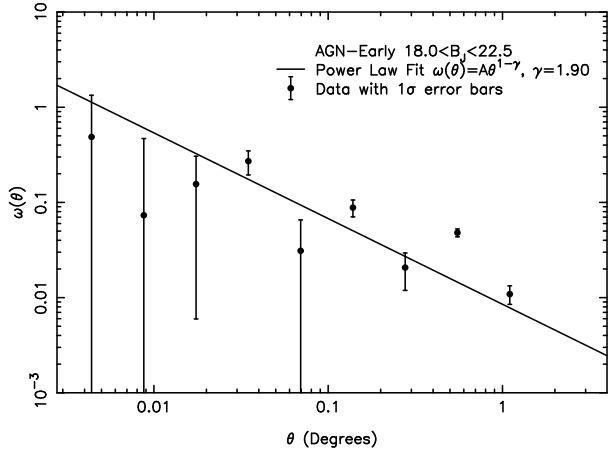}}
\centerline{\psfig{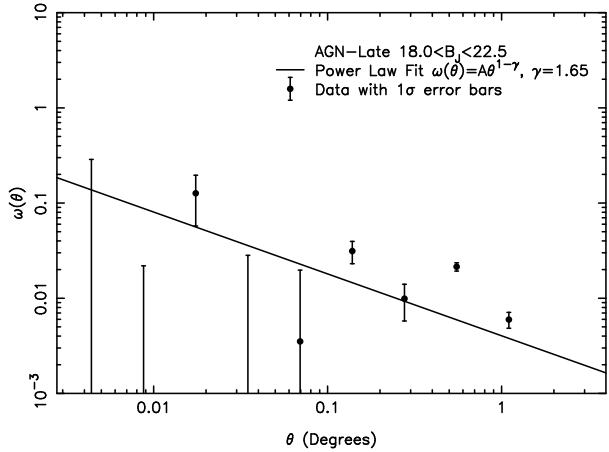}}
\figcaption[MJIBrown.fig9a.ps MJIBrown.fig9b.ps]
{The AGN-early and AGN-late angular cross-correlation functions. The 
AGN-early correlation function is significantly 
stronger than the AGN-late correlation
function. While a power-law is a good approximation to the observed AGN-early
cross-correlation function, a power-law fit to the AGN-late 
cross-correlation function is a poor fit to the data. 
\label{fig:qsogal}}
\end{figure}

The AGN-early and AGN-late angular cross-correlation functions for 
$B_J<22.5$ galaxies are shown in Figure \ref{fig:qsogal}.
Power-laws with $\gamma$ fixed to $1.90$ and $1.65$ have been 
fitted to the AGN-early and AGN-late data on angular scales 
$<0.3^\circ$. The AGN-early cross-correlation function is 
strong and the power-law fit is a good approximation 
to the observed clustering though most of the data points are
$\simgt 1\sigma$ from the fit. In contrast, the AGN-late cross-correlation
is weak and a power-law is a poor fit to the data. From the
available data, the amplitude of the AGN-late cross-correlation
function can not be reliably measured. 

\begin{figure}
\centerline{\psfig{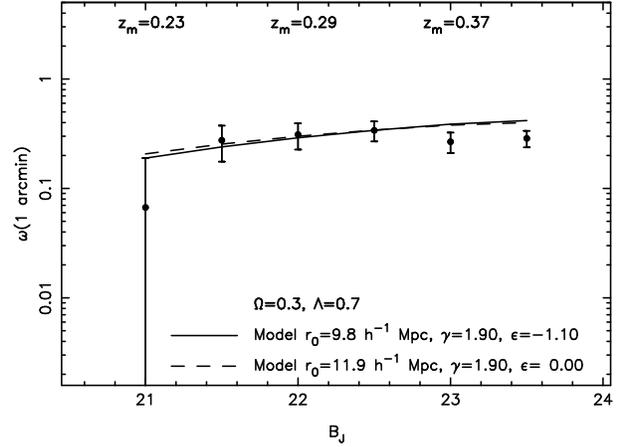}}
\figcaption[MJIBrown.fig10.ps]
{The amplitude of the AGN-early angular cross-correlation
function. 
Models with clustering fixed in comoving and physical coordinates
have been fitted to the $B_J=22.5$ data point. The clustering is 
significantly stronger than the autocorrelation function of early subsample 
galaxies.
\label{fig:qeampmag}}
\end{figure}

The amplitude of the AGN-early cross-correlation function is plotted in 
Figure \ref{fig:qeampmag} along with models of the cross-correlation
with evolution fixed in comoving and physical coordinates. 
While the amplitude of the angular cross-correlation function is 
typically less than the early subsample autocorrelation function, the difference
between the early subsample and AGN redshift distributions results
in the estimates of $r_0$ being significantly higher than that of
the early subsample autocorrelation function. 

\begin{figure}
\centerline{\psfig{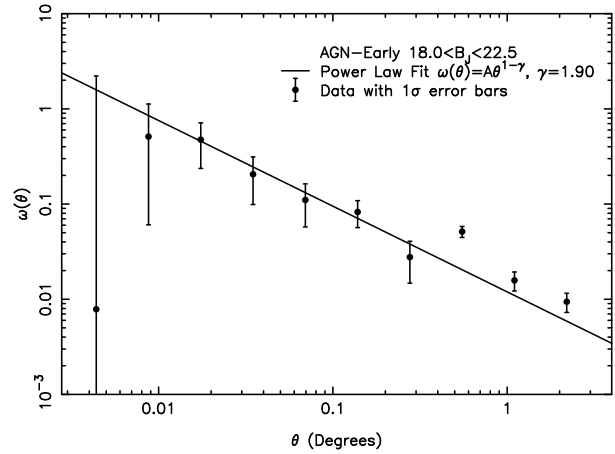}}
\figcaption[MJIBrown.fig11.ps]
{The angular cross-correlation of AGN and $B_J<22.5$ early
subsample galaxies with photometric redshift constraints applied to the 
pair counts. Poisson statistics have been used 
to determine the $1\sigma$ errors shown with the data points. A power
law with $\gamma=1.90$ is a good fit to the data.
\label{fig:photoqe}}
\end{figure}

The errors of the $\omega(\theta)$ estimates are dominated 
by galaxies along the line-of-sight which 
are not associated with the AGN. The signal and significance of the 
correlation function can be improved if the number of unassociated galaxies
included in the estimate of $\omega(\theta)$ are reduced. 
Early-type galaxies have comparatively small photometric redshift errors
so it is possible to define a narrow range of photometric redshifts which
could be associated with the AGN. Figure \ref{fig:photoqe} 
shows the AGN-early cross-correlation function for galaxies with 
photometric redshifts within $2\sigma$ of the AGN redshift. The 
inclusion of photometric redshift constraints has significantly 
improved the signal-to-noise and the observed correlation function
is more consistent with a power-law than Figure \ref{fig:qsogal}.
While deviations from a power-law have been seen in 
some estimates of AGN-galaxy correlation functions \citep{hal98, cro99}, 
there are no  statistically significant breaks from a power-law on 
scales $<0.3^\circ$ in Figure \ref{fig:photoqe}.

\begin{figure}
\centerline{\psfig{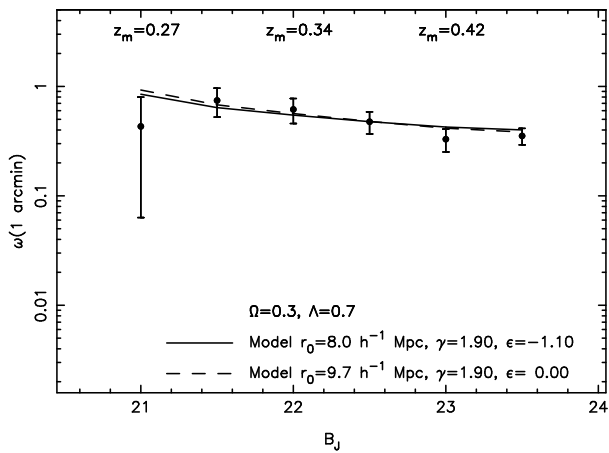}}
\figcaption[MJIBrown.fig12.ps]
{The amplitude of the AGN-early cross-correlation function with 
photometric redshift constraints applied. The errors are significantly
reduced and the best fits to the data are slightly decreased. However,
the amplitude of the clustering is consistent with AGN being in 
environments comparable to elliptical galaxies. 
The estimated median redshifts are for early subsample galaxies
that satisfy the photometric redshift constraints rather than the 
entire early subsample.
\label{fig:pqeampmag}}
\end{figure}

The spatial AGN-early cross-correlation can be determined
by deriving the redshift distribution of galaxies with photometric redshifts 
within $2\sigma$ of each AGN when applying Limber's equation. The amplitude
of the AGN-early cross-correlation and models of the spatial 
cross-correlation are shown in Figure \ref{fig:pqeampmag} and 
the fit to the $B_J=22.5$ data point is weaker than in 
Figure \ref{fig:qeampmag}. However, the estimate of $r_0$ and 
the power-law form of the angular correlation function in 
Figure \ref{fig:photoqe} are consistent with AGN environments
being similar to the environments of strongly clustered galaxies such as the 
early subsample and low redshift ellipticals \citep{guz97}.

The estimates of $r_0$ are dependent on the assumed value of
$\gamma$ and it is possible that $\gamma\ll 1.90$. If radio-quiet
AGN were in spiral galaxies \citep{hut84, mal84} 
and had similar clustering properties to late subsample galaxies, the value of 
$\gamma$ for the AGN-early cross-correlation function 
would be $\sim 1.65$. However, when the AGN-early cross-correlation
function was fitted with power-laws with $\gamma\leq 1.65$, 
the estimates of $r_0$ increased to $> 11 h^{-1} {\rm Mpc}$ which 
is approximately double the estimate of $r_0$ for the early-late 
spatial cross-correlation function. 

\begin{figure}
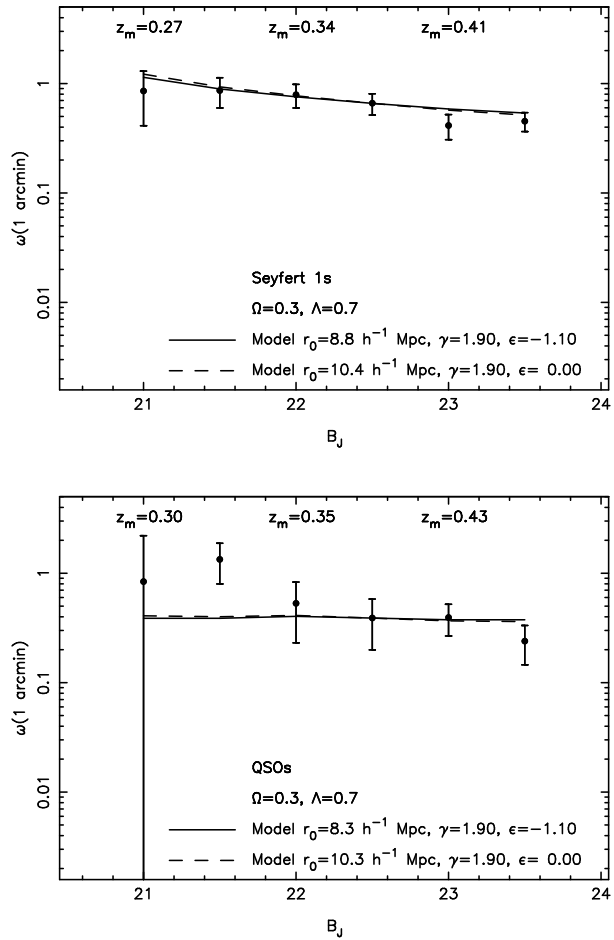

\centerline{\psfig{file=MJIBrown.fig13a.ps,angle=270,width=8.0cm}}
\centerline{\psfig{file=MJIBrown.fig13b.ps,angle=270,width=8.0cm}}
\figcaption[MJIBrown.fig13a.ps MJIBrown.fig13b.ps]
{The amplitude of the Seyfert 1 and QSO cross-correlation
functions with photometric redshift constraints. 
There is no evidence for a strong correlation between 
AGN luminosity and host environment.
\label{fig:bf}}
\end{figure}

There may be a correlation between AGN luminosity and host galaxy 
if luminous AGN occur more frequently in early-type galaxies 
than Seyferts \citep{mcl99}. If this is the case,
the correlation between galaxy morphology and environment may 
result in a correlation between AGN luminosity and environment.
Figure \ref{fig:bf} compares the amplitudes of the 
Seyfert 1 ($M_{B_J}<-21.5$) and QSO ($M_{B_J}>-21.5$) cross-correlation 
functions with the early subsample. 
Fits to the $B_J=22.5$ data point show Seyfert 1s are correlated with 
slightly richer environments but difference between the $r_0$ estimates
has $< 1\sigma$ significance. 

\section{Discussion}

While the detection of radio-quiet AGN in environments similar to 
early-type galaxies may appear to contradict previous 
studies at similar redshifts \citep{yee87, boy93, cro99, tep99, smi00}, this assumes 
all studies of AGN-galaxy clustering have used the same galaxy population
to measure AGN environments. However, as color selection 
has been used to select galaxies for 
this study, the properties of the early and late subsamples will
differ significantly from catalogues of galaxies which are 
selected by observed broadband flux. 

Early subsample galaxies contribute $\sim 30\% $ of the 
total galaxy counts at $B_J<21$ and only $\sim 15\%$ of the counts
at $B_J>22$. If the AGN-late cross-correlation function is assumed to 
be negligible and the AGN-early cross-correlation function is a power
law with $r_0\sim 9 h^{-1} {\rm Mpc}$, then the AGN-galaxy 
cross-correlation function should be a power-law with 
$r_0 \sim 5 h^{-1} {\rm Mpc} $ at $B_J<21$ and 
$r_0 \sim 3 h^{-1} {\rm Mpc} $ at $B_J>22$. In redder bands
where the early-type galaxies comprise a larger fraction of the 
total galaxy number counts, stronger clustering will be measured. 
Early subsample galaxies comprise $\sim 25\%$ of the 
total $R<21.5$ galaxy counts and the AGN-galaxy correlation function 
would be expected to be $r_0 \sim 4 h^{-1} {\rm Mpc}$ even 
though the redshift range sampled is comparable to that of $B_J\sim 23$
galaxies. These estimates of $r_0$ for $B_J$ and $R$ limited samples
are consistent with measurements of the AGN-galaxy cross-correlation 
function derived from single band imaging by \cite{yee87}, \cite{ell91} and \cite{smi95, smi00}. 

While radio-quiet AGN occur in environments comparable to early-type
galaxies, this should not be interpreted as a direct correlation 
between richness and AGN activity. As shown in Figure \ref{fig:bf}, 
there is no evidence for a strong correlation between AGN 
luminosity and environment. HST imaging of radio-quiet QSOs and 
X-ray selected AGN indicates $\sim 75\%$ of host galaxies have 
morphologies earlier than Sbc galaxies \citep{mcl99, sch00}. 
It is therefore plausible that AGN activity is not significantly effected by 
the host environment and that the observed strength of the AGN-early
cross-correlation function is due to the correlation between 
galaxy morphology and clustering properties. 

The strength AGN-early cross-correlation function implies that 
it should be detectable at $z\gg 0.5$. Using samples of 
early-type galaxies obtained with CCD mosaics on 4-m class telescopes, it
will be possible to measure the evolution of the AGN-early cross-correlation 
function at $z>1$. This will provide an estimate of the bias of AGN with 
respect to galaxies which are used to measure large-scale-structure at lower 
redshifts. If the clustering properties of AGN are due to the distribution 
of host galaxy morphologies, the evolution of the AGN-early cross-correlation 
function will also place strong constraints on the properties
of AGN host galaxies as a function of redshift. 

\section{Summary}

The environments of galaxies and 69 $0.2<z<0.7$ AGN have been 
measured using photometric redshifts and color criteria to 
select galaxy types to $B_J\sim 23.5$. The key conclusions
are \\
\noindent
(i) The clustering of early subsample galaxies is strong across the 
observed magnitude range with $r_0\sim 7 h^{-1} {\rm Mpc}$ and
$\gamma\sim 1.90$. \\
(ii) The autocorrelation function of late subsample galaxies is weak
and decreases to $r_0\simlt 4 h^{-1} {\rm Mpc}$ at $B_J\sim 23.5$. This
is probably due to the increasing fraction of weakly clustered blue
galaxies at faint apparent magnitudes.\\
(iii) The cross-correlation function of radio-quiet $-24<M_B<-19$ AGN 
with early subsample galaxies has been detected with high significance on 
scales $\simlt 1^\circ$. The AGN-early spatial cross-correlation function 
is stronger than the early subsample autocorrelation function and 
is comparable to the clustering of elliptical galaxies at $z\sim 0$.\\
(iv) The AGN-late cross-correlation function is very weak and has
been detected with low significance. 
As the fraction of late-type galaxies in magnitude limited
samples increases with survey depth, the cross-correlation function 
of AGN with all galaxies decreases with increasing magnitude.\\
(v) The signal-to-noise of AGN-galaxy cross-correlation functions
is significantly improved by using photometric redshifts to 
reject galaxies which can not be associated with the AGN from the 
correlation function estimate.\\
(vi) The correlation between AGN optical luminosity and host environment is weak
and has not been detected at a significant level in this work.\\

\section{Acknowledgments}

The authors wish to thank the SuperCOSMOS unit at Royal Observatory
Edinburgh for providing the digitized scans of UK Schmidt
photographic plates. The authors also wish to thank Nigel Hambly, Bryn Jones
and Harvey MacGillivray for productive discussions of
the methods employed to coadd scans of photographic plates.
This research has made use of the NASA/IPAC Extragalactic Database
which is operated by the Jet Propulsion
Laboratory, California Institute of Technology, under contract with the
National Aeronautics and Space Administration. Michael Brown
acknowledges the financial support of an Australian Postgraduate Award
and the support of the National Optical Astronomy Observatory
which is operated by the Association of Universities for 
Research in Astronomy, Inc., under a cooperative agreement 
with the National Science Foundation.

\clearpage

\begin{deluxetable}{llccccc}
\setlength{\tabcolsep}{0.02in} 
\rotate
\tabletypesize{\scriptsize}
\tablecaption{A sample of studies of AGN environments\label{table:qsogal}}
\tablehead{
Reference & AGN 	& Number & AGN      & AGN	& Galaxy    & Correlation \\
          & Type         & of    &  Redshift & Magnitudes & Magnitude & Function     \\
          &             & AGN   &  Range    & 	 	& Limit     & Estimate   \\
}
\startdata
\cite{der98}       	& Seyferts
& 33    & $z<0.04$     	& $-23.0<M_R<-18.5$ & $M_R<-19.5$   & $r_0 \sim 5.4 h^{-1} {\rm Mpc}$ \\
\cite{lau95}     & Seyfert 1
& 55 &   $z<0.05$        & $-22.0<M_V< -17.0$ & $B_J<21$      & $r_0\sim 5 h^{-1} {\rm Mpc}$ \\
\cite{lau95}     & Seyfert 2
& 49    & $z<0.05$      &  $-21.5<M_V<-21.5$ & $B_J<21$      & $r_0> 5 h^{-1} {\rm Mpc}$ \\
\cite{yee87}      & RQ QSOs
& 33    & $z<0.3$   & $-26.0\simlt M_r \simlt -23.5$ & $r\simlt 21.5$ & $r_0\sim 8 h^{-1} {\rm Mpc}$ \\
\cite{smi95}    & X-ray QSOs
&  169  & $z<0.3$       & $-23.5<M_V<-17.0$ & $B_J<20.5$ & $r_0\sim 5 h^{-1} {\rm Mpc}$ \\
\cite{yee87}      & RQ QSOs
& 7    & $0.3<z<0.6$   & $-26.5 \simlt M_r \simlt -23.5$ & $r\simlt 21.5$ & $r_0\sim 5 h^{-1} {\rm Mpc}$ \\
\cite{ell91}   & RQ QSOs
& 46 & $0.3<z<0.6$      & $-24.0<M_R<-18.5$ & $r\sim23.5$ & $r_0\sim 6 h^{-1} {\rm Mpc}$
\\
This work (AGN-Early)   & UBR AGN
& 69    & $0.2<z<0.7$   & $-24.0<M_B<-19.0$ & $B_J<23.5$    & $r_0\sim 8 h^{-1} {\rm Mpc}$\\
\cite{smi00} & X-ray QSOs
& 83 & $0.3<z<0.7$      & $-26.0<M_V<-17.5$ & $V=23$        & $r_0\sim 3h^{-1} {\rm Mpc} $ \\
\cite{boy93}            & QSOs
& 27    & $0.9<z<1.7$   & $-24.0<M_R<-22.0$ & $R<23$        & No correlation \\
\cite{tep99}  & RQ QSOs
& 30    & $0.9<z<2.1$   & $-26.5<M_V<-20.6$ & $I,J,H\sim 21,K$ & $2\sigma$ correlation \\
\cite{cro99}          & RQ QSOs
& 150   & $0.0<z<3.2$   & $B\simlt 22$ & $B_J<23$      & No correlation \\
\\
\cite{yee87}      & RL QSOs
& 10    & $0.3<z<0.5$   & $-25.0 \simlt M_r \simlt -23.5$ & $r\simlt 21.5$ & $r_0\sim 9 h^{-1} {\rm Mpc}$ \\
\cite{yee87}      & RL QSOs
& 9     & $0.55<z<0.65$ & $-25.5 \simlt M_r \simlt -23.5$ & $r\simlt 21.5$ & $r_0\sim 17 h^{-1} {\rm Mpc}$ \\
\cite{wol00}  & RL QSOs
& 21    & $0.50<z<0.82$ & $-24.3<M_B<-19.9$ & $V,R\sim 23.5,I$ & $r_0\sim 11.7 h^{-1} {\rm Mpc}$ \\\cite{hal98}     & RL QSOs
& 31    & $1.0<z<2.0$   & $-27.5<M_V<-24.0$ & $K\simgt 19$  & Rich. $\sim 0$ clusters\\
\enddata
\end{deluxetable}

\clearpage

\begin{deluxetable}{lrcrrcr}
\tablecaption{The Early Subsample Angular Correlation Function\label{table:ee}}
\tablehead{
Field & \multicolumn{3}{c}{SGP} &  \multicolumn{3}{c}{F855} \\
Magnitude Range & $N_{gal}$ & $\gamma$   & $\omega(1^\prime) \times 10^3$       &
 $N_{gal}$ & $\gamma$    & $\omega(1^\prime) \times 10^3$  \\
}
\startdata
$18.0 \leq B_J \leq 20.0$ & 2206  & $1.70\pm 0.11$ & $1352 \pm 233$    
                          & 2599  & $1.79\pm 0.31$ & $782 \pm 196$ \\
$18.0 \leq B_J \leq 21.0$ & 5187  & $1.79\pm 0.08$ & $830 \pm 83$    
                          & 5553  & $2.02\pm 0.19$ & $525 \pm 78$ \\
$18.0 \leq B_J \leq 22.0$ & 11056  & $1.74\pm 0.04$ & $510 \pm 51$    
                          & 11303  & $1.94\pm 0.09$ & $326 \pm 34$ \\
$18.0 \leq B_J \leq 23.0$ & 21833 & $1.81\pm 0.05$ & $283 \pm 28$    
                          & 21652 & $1.87\pm 0.07$ & $194 \pm 19$ \\
$18.0 \leq B_J \leq 23.5$ & 29290 & $1.84\pm 0.06$ & $221 \pm 22$    
                          & 27915 & $1.89\pm 0.05$ & $169 \pm 17$ \\
\enddata
\end{deluxetable}

\begin{deluxetable}{lrcrrcr}
\tablecaption{The Late Subsample Angular Correlation Function\label{table:ll}}
\tablehead{
Field & \multicolumn{3}{c}{SGP} &  \multicolumn{3}{c}{F855} \\
Magnitude Range & $N_{gal}$ & $\gamma$   & $\omega(1^\prime) \times 10^3$       &
 $N_{gal}$ & $\gamma$    & $\omega(1^\prime) \times 10^3$  \\
}
\startdata
$18.0 \leq B_J \leq 20.0$ & 3854   	& $1.35\pm 0.27$ & $443 \pm 257$    
                          & 4648	& $1.57\pm 0.37$ & $289 \pm 87$ \\
$18.0 \leq B_J \leq 21.0$ & 11952  	& $1.39\pm 0.13$ & $269 \pm 25$    
                          & 13866  	& $1.62\pm 0.16$ & $157 \pm 21$ \\
$18.0 \leq B_J \leq 22.0$ & 37785  	& $1.33\pm 0.10$ & $120 \pm 9$    
                          & 43047 	& $1.42\pm 0.15$ & $76 \pm 7$ \\
$18.0 \leq B_J \leq 23.0$ & 128278 	& $1.29\pm 0.09$ & $44 \pm 8$    
                          & 134331	& $1.31\pm 0.18$ & $37 \pm 16$ \\
$18.0 \leq B_J \leq 23.5$ & 207390 	& $1.34\pm 0.12$ & $28 \pm 4$    
                          & 183347 	& $1.33\pm 0.12$ & $27 \pm 5$ \\
\enddata
\end{deluxetable}

\begin{deluxetable}{lcrcr}
\tablecaption{The Early-Late Angular Cross-Correlation Function\label{table:el}}
\tablehead{
Field & \multicolumn{2}{c}{SGP} &  \multicolumn{2}{c}{F855} \\
Magnitude Range & $\gamma$   & $\omega(1^\prime) \times 10^3$ 
& $\gamma$    & $\omega(1^\prime) \times 10^3$  \\
}
\startdata
$18.0 \leq B_J \leq 20.0$ & $1.43\pm 0.25$ & $430 \pm 105$    
                          & $1.33\pm 0.35$ & $202 \pm 70$ \\
$18.0 \leq B_J \leq 21.0$ & $1.67\pm 0.08$ & $322 \pm 30$    
                          & $1.48\pm 0.14$ & $204 \pm 22$ \\
$18.0 \leq B_J \leq 22.0$ & $1.73\pm 0.06$ & $183 \pm 10$    
                          & $1.58\pm 0.08$ & $102 \pm 8$ \\
$18.0 \leq B_J \leq 23.0$ & $1.76\pm 0.04$ & $85  \pm 4$    
                          & $1.59\pm 0.08$ & $52  \pm 3$ \\
$18.0 \leq B_J \leq 23.5$ & $1.65\pm 0.04$ & $56  \pm 3$    
                          & $1.46\pm 0.09$ & $40  \pm 2$ \\
\enddata
\end{deluxetable}

\begin{deluxetable}{cccccc}
\tablecaption{The spatial autocorrelation and cross-correlation
functions of galaxies.\label{table:r0}}
\tablehead{
Correlation & Galaxy Magnitude & Field & $\gamma$ &
$r_0$ ($h^{-1} {\rm Mpc}$)              & $r_0$ ($h^{-1} {\rm Mpc}$)  \\
Function Samples & Range          &          &          &
$\epsilon=1-\gamma$     & $\epsilon=0$ }
\startdata
Early-Early & $18.0<B_J<22.5$ & SGP & 1.90
& $6.6\pm 0.3$ & $7.7\pm 0.3$ \\
Early-Early & $18.0<B_J<22.5$ & F855 & 1.90
& $5.3\pm 0.3$ & $6.1\pm 0.3$ \\
Late-Late  & $18.0<B_J<22.5$ & SGP & 1.35
& $4.5\pm 0.2$ & $5.9\pm 0.3$ \\
Late-Late  & $18.0<B_J<22.5$ & F855 & 1.35
& $3.5\pm 0.2$ & $4.7\pm 0.3$ \\
Early-Late  & $18.0<B_J<22.5$ & SGP & 1.65
& $5.0\pm 0.2$ & $6.1\pm 0.2$ \\
Early-Late  & $18.0<B_J<22.5$ & F855 & 1.65
& $3.8\pm 0.2$ & $4.7\pm 0.2$ \\
\\
Early-Early & $18.0<B_J<23.5$ & SGP & 1.90
& $5.8\pm 0.2$ & $7.0\pm 0.2$ \\
Early-Early & $18.0<B_J<23.5$ & F855 & 1.90
& $5.2\pm 0.3$ & $6.2\pm 0.3$ \\
Late-Late  & $18.0<B_J<23.5$ & SGP & 1.35
& $2.5\pm 0.2$ & $3.6\pm 0.3$ \\
Late-Late  & $18.0<B_J<23.5$ & F855 & 1.35
& $2.5\pm 0.2$ & $3.5\pm 0.3$ \\
Early-Late  & $18.0<B_J<23.5$ & SGP & 1.65
& $4.0\pm 0.1$ & $5.1\pm 0.2$ \\
Early-Late  & $18.0<B_J<23.5$ & F855 & 1.65
& $3.0\pm 0.2$ & $4.0\pm 0.2$ \\
\enddata
\end{deluxetable}

\begin{deluxetable}{lcrrr}
\tablecaption{The cross-correlation function of $0.2<z<0.7$ AGN and early 
subsample galaxies. The values of $\hat\omega(\theta)$ have been corrected
for the integral constraint which is $\sim 10\%$ of the estimate
of the correlation function at $1^\prime$.\label{table:qe}}
\tablehead{
\multicolumn{5}{c}{AGN-Early}\\
Magnitude Range & Angle & $AG$ & $AR\times 20$ & 
\multicolumn{1}{c}{$\hat\omega(\theta)$} \\
}
\startdata
$18.0 \leq B_J \leq 22.5$ 
& $10^{\prime\prime}\leq \theta <20^{\prime\prime}$ &
6 & 81 		& 	$0.48\pm 0.84$ 	\\
$18.0 \leq B_J \leq 22.5$ 
& $20^{\prime\prime}\leq \theta <40^{\prime\prime}$ &
15 & 281  	& 	$0.07\pm 0.39$ 	\\
$18.0 \leq B_J \leq 22.5$ 
& $40^{\prime\prime}\leq \theta <80^{\prime\prime}$ &
59 & 1026 	&	$0.15\pm 0.15$ 	\\
$18.0 \leq B_J \leq 22.5$ 
& $80^{\prime\prime}\leq \theta <160^{\prime\prime}$ &
272 & 4299 	&	$0.27\pm 0.07$	\\
$18.0 \leq B_J \leq 22.5$ 
& $160^{\prime\prime}\leq \theta <320^{\prime\prime}$ &
882 & 17203 	&   	$0.03\pm 0.03$  	\\
$18.0 \leq B_J \leq 22.5$ 
& $320^{\prime\prime}\leq \theta <640^{\prime\prime}$ &
3704 & 68428 	&   	$0.08\pm 0.02$  	\\
\\
$18.0 \leq B_J \leq 23.5$ 
& $10^{\prime\prime}\leq \theta <20^{\prime\prime}$ &
10 & 132 & 	$0.52 \pm 0.61$\\
$18.0 \leq B_J \leq 23.5$ 
& $20^{\prime\prime}\leq \theta <40^{\prime\prime}$ &
28 & 520 & 	$0.08 \pm 0.21$\\
$18.0 \leq B_J \leq 23.5$ 
& $40^{\prime\prime}\leq \theta <80^{\prime\prime}$ &
114 & 1956 &	$0.17 \pm 0.11$\\
$18.0 \leq B_J \leq 23.5$ 
& $80^{\prime\prime}\leq \theta <160^{\prime\prime}$ &
497 & 8128 &	$0.04 \pm 0.06$\\
$18.0 \leq B_J \leq 23.5$ 
& $160^{\prime\prime}\leq \theta <320^{\prime\prime}$ &
1676 & 32371 &  $0.07 \pm 0.03$ \\
$18.0 \leq B_J \leq 23.5$ 
& $320^{\prime\prime}\leq \theta <640^{\prime\prime}$ &
6881 & 129220 & $0.02\pm 0.01$  	\\
\enddata
\end{deluxetable}

\begin{deluxetable}{lcrrr}
\tablecaption{The cross-correlation function of $0.2<z<0.7$ AGN and late
subsample galaxies. 
\label{table:ql}}
\tablehead{
\multicolumn{5}{c}{AGN-Late}\\
Magnitude Range & Angle & $AG$ & $AR\times 20$ & 
\multicolumn{1}{c}{$\hat\omega(\theta)$} \\
}
\startdata
$18.0 \leq B_J \leq 22.5$ 
& $10^{\prime\prime}\leq \theta <20^{\prime\prime}$ 
& 15 & 320	& 	$-0.06\pm 0.35$ \\
$18.0 \leq B_J \leq 22.5$ 
& $20^{\prime\prime}\leq \theta <40^{\prime\prime}$ 
& 53 & 1183	& 	$-0.10\pm 0.12$ \\
$18.0 \leq B_J \leq 22.5$ 
& $40^{\prime\prime}\leq \theta <80^{\prime\prime}$ 
& 263 & 4680	& 	$0.13\pm 0.07$ \\
$18.0 \leq B_J \leq 22.5$ 
& $80^{\prime\prime}\leq \theta <160^{\prime\prime}$ 
& 949 & 19111 	& 	$0.00\pm 0.03$\\
$18.0 \leq B_J \leq 22.5$ 
& $160^{\prime\prime}\leq \theta <320^{\prime\prime}$ 
& 3812 & 76187  & 	$0.00\pm 0.02$ \\ 
$18.0 \leq B_J \leq 22.5$ 
& $320^{\prime\prime}\leq \theta <640^{\prime\prime}$ 
& 15602 & 303404 &   	$0.03\pm 0.01$  	\\
\\
$18.0 \leq B_J \leq 23.5$ 
& $10^{\prime\prime}\leq \theta <20^{\prime\prime}$ 
& 42 & 905	& $-0.07 \pm 0.14$ \\
$18.0 \leq B_J \leq 23.5$ 
& $20^{\prime\prime}\leq \theta <40^{\prime\prime}$ 
& 175 & 3333	& $0.05 \pm 0.08$ \\
$18.0 \leq B_J \leq 23.5$ 
& $40^{\prime\prime}\leq \theta <80^{\prime\prime}$ 
& 751 & 13747	& $0.09 \pm 0.04$ \\
$18.0 \leq B_J \leq 23.5$ 
& $80^{\prime\prime}\leq \theta <160^{\prime\prime}$ 
& 2805 & 55104	& $0.02 \pm 0.02$ \\
$18.0 \leq B_J \leq 23.5$ 
& $160^{\prime\prime}\leq \theta <320^{\prime\prime}$ 
& 11166 & 221132 & $0.01 \pm 0.01$ \\ 
$18.0 \leq B_J \leq 23.5$ 
& $320^{\prime\prime}\leq \theta <640^{\prime\prime}$ &
45032 & 880067 	& $0.03\pm 0.01$  	\\
\enddata
\end{deluxetable}

\begin{deluxetable}{lcrrr}
\tablecaption{The cross-correlation of AGN with the early subsample with 
photometric redshift constraints applied to the data.\label{table:pqe}}
\tablehead{Magnitude Range & 
Angle & 
$AG$ & 
$AR\times 20$ & 
$\hat\omega(\theta)$ \\
}
\startdata
$18.0 \leq B_J \leq 22.5$ & $10^{\prime\prime}\leq \theta <20^{\prime\prime}$ &
1 & 20 & $0.07\pm 2.20$ \\
$18.0 \leq B_J \leq 22.5$ & $20^{\prime\prime}\leq \theta <40^{\prime\prime}$ &
10 & 133 & $0.51\pm0.61$ \\
$18.0 \leq B_J \leq 22.5$ & $40^{\prime\prime}\leq \theta <80^{\prime\prime}$ &
38 & 518 & $0.48\pm0.23$ \\
$18.0 \leq B_J \leq 22.5$ & $80^{\prime\prime}\leq \theta <160^{\prime\prime}$ &
126 & 2104 & $0.21\pm0.10$\\
$18.0 \leq B_J \leq 22.5$ & $160^{\prime\prime}\leq \theta <320^{\prime\prime}$ &
438 & 7947 & $0.11\pm0.05$ \\
$18.0 \leq B_J \leq 22.5$ & $320^{\prime\prime}\leq \theta <640^{\prime\prime}$ &
1704 & 31708 & $0.08\pm0.03$ \\
\\
$18.0 \leq B_J \leq 23.5$ & $10^{\prime\prime}\leq \theta <20^{\prime\prime}$ &
5 & 71 & $0.41\pm0.91$ \\
$18.0 \leq B_J \leq 23.5$ & $20^{\prime\prime}\leq \theta <40^{\prime\prime}$ &
23 & 373 & $0.24\pm0.26$ \\
$18.0 \leq B_J \leq 23.5$ & $40^{\prime\prime}\leq \theta <80^{\prime\prime}$ &
89 & 1388 & $0.28\pm0.13$ \\
$18.0 \leq B_J \leq 23.5$ & $80^{\prime\prime}\leq \theta <160^{\prime\prime}$ &
340 & 5637 & $0.21\pm0.07$\\
$18.0 \leq B_J \leq 23.5$ & $160^{\prime\prime}\leq \theta <320^{\prime\prime}$ &
1172 & 21818 & $0.08\pm0.03$ \\
$18.0 \leq B_J \leq 23.5$ & $320^{\prime\prime}\leq \theta <640^{\prime\prime}$ &
4661 & 87823 & $0.08\pm0.02$ \\
\enddata
\end{deluxetable}

\end{document}